\documentstyle [12pt,epsfig] {article}

\catcode`@=12
\begin{document}
\thispagestyle{empty}
\begin{flushright} UCRHEP-T303\\March 2001\
\end{flushright}
\vspace{0.5in}
\begin{center}
{\Large  COLLIDER VERIFICATION OF THE NEUTRINO\\ MASS MATRIX 
 IN TWO SCENARIOS\\}
\vspace{1.2in}
{\bf Ernest Ma\\}
\vspace{0.2in}
{\sl Department of Physics, University of California, 
Riverside, CA 92521, USA} 
\vspace{1.2in}
\end{center}
\begin{abstract}\
If the origin of neutrino mass is at the TeV energy scale, 
collider experiments may in fact {\it map out} all the elements of the 
$3 \times 3$ neutrino mass matrix, up to an overall scale.  Two examples
\cite{mrs,ma01} are discussed and one is related \cite{mr} to the muon 
anomalous magnetic moment.
\end{abstract}
\vspace{0.1in}
--------------

\noindent Talk given at the 9th International Workshop on ``Neutrino 
Telescopes'' (Venice, Italy) March 6-9, 2001.

\newpage
\baselineskip 24pt

\section{Introduction}

In the minimal Standard Model with one Higgs doublet $\Phi = (\phi^+,\phi^0)$ 
and 3 lepton doublets $L = (\nu, l)_L$ and singlets $l_R$ only, neutrino 
mass must come from the effective dimension-5 operator \cite{wein,ma98}
\begin{equation}
{1 \over \Lambda} L L \Phi \Phi = {1 \over \Lambda} (\nu \phi^0 - l \phi^+)^2,
\end{equation}
which shows that the form of $m_\nu$ must necessarily be ``seesaw'', i.e. 
$v^2 / \Lambda$ where $v = \langle \phi^0 \rangle$, whatever the underlying 
mechanism for neutrino mass is.

The canonical seesaw mechanism \cite{seesaw} assumes 3 heavy right-handed 
singlet lepton fields $N_R$ with the Yukawa couplings $f L N \Phi$ and the 
Majorana mass $m_N$, hence Eq.(1) is realized with the famous expression
\begin{equation}
m_\nu = {f^2 v^2 \over m_N}.
\end{equation}
Note that lepton number is violated by $m_N$ in the denominator and it should 
be large for a small neutrino mass, i.e.
\begin{equation}
m_\nu \sim \left( {f \over 1.0} \right)^2 \left( {10^{13}~{\rm GeV} \over m_N} 
\right) ~{\rm eV}.
\end{equation}

\section{Higgs Triplet Model}

An equally satisfactory realization of Eq.(1) is to use a Higgs triplet 
\cite{sv,ms} $\xi = (\xi^{++},\xi^+,\xi^0)$ with
\begin{eqnarray}
{\cal L}_{int} &=& f_{ij} [\xi^0 \nu_i \nu_j + \xi^+ (\nu_i l_j + l_i \nu_j)/
\sqrt 2 + \xi^{++} l_i l_j] \nonumber \\ &+& \mu (\bar \xi^0 \phi^0 \phi^0 
- \sqrt 2 \xi^- \phi^+ \phi^0 + \xi^{--} \phi^+ \phi^+) + h.c.
\end{eqnarray}
Instead of having a negative $m_\xi^2$, make it positive and large, i.e. 
$m_\xi >> v$. We then find \cite{ms}
\begin{equation}
m_\nu = {2 f_{ij} \mu v^2 \over m_\xi^2} = 2 f_{ij} \langle \xi^0 \rangle.
\end{equation}
Note that the effective operator of Eq.(1) is realized here with a simple 
rearrangement of the individual terms, i.e.
\begin{equation}
L_i L_j \Phi \Phi = \nu_i \nu_j \phi^0 \phi^0 - (\nu_i l_j + l_i \nu_j) \phi^+ 
\phi^0 + l_i l_j \phi^+ \phi^+.
\end{equation}
 
Note also that lepton number is violated in the numerator in this case. 
If $f_{ij} \sim 1$, then $\mu/m_\xi^2 < 10^{-13}$ GeV$^{-1}$.  Hence 
$m_\xi \sim 1$ TeV is possible, if $\mu < 100$ eV.  To obtain such a 
small mass parameter, the ``shining'' mechanism of extra large dimensions 
\cite{ahd} may be used. \cite{mrs} Let $\chi$ be a singlet scalar in the bulk 
carrying lepton number $L = -2$, then
\begin{equation}
\langle \chi \rangle \sim {\Gamma ({n-2 \over 2}) \over 4 \pi^{n \over 2}} 
M_* \left( M_* \over M_P \right)^{2 - {4 \over n}}.
\end{equation}
For $n=3$, $M_* \sim 1$ TeV, $M_P = 2.4 \times 10^{18}$ GeV, $\langle \chi 
\rangle \sim 4.4$ eV.  Therefore, if we replace $\mu$ by $h \chi$, the 
intriguing possibility of having $m_\xi \sim 1$ TeV is realized.  In 
particular, the doubly charged $\xi^{\pm \pm}$ can be easily produced at 
colliders and $\xi^{++} \to l_i^+ l_j^+$ is a distinct and backgroundless 
decay which maps out $|f_{ij}|$, and thus determine directly the neutrino 
mass matrix up to an overall scale. \cite{mrs}

\section{Leptonic Higgs Doublet Model}

Another simple and interesting way to have the origin of neutrino mass at 
the TeV scale has just been proposed. \cite{ma01}  As in the canonical 
seesaw model, we have again 3 $N_R$'s but they are now assigned $L=0$ 
instead of the customary $L=1$.  Hence the Majorana mass terms are 
allowed but the usual $L N \Phi$ terms are forbidden by lepton-number 
conservation.  The $LL\Phi\Phi$ operator of Eq.(1) is not possible and 
$m_\nu = 0$ at this point.

We now add a new scalar doublet $\eta = (\eta^+,\eta^0)$ with $L=-1$, then 
$f L N \eta$ is allowed, and the operator $LL\eta\eta$ will generate a 
nonzero neutrino mass if $\langle \eta^0 \rangle \neq 0$.  The trick now 
is to show how $f \langle \eta^0 \rangle < 1$ MeV can be obtained naturally, 
so that $m_N \sim 1$ TeV becomes possible and amenable to experimental 
verification, in contrast to the very heavy $N_R$'s of the canonical seesaw 
mechanism.

Consider the following Higgs potential:
\begin{eqnarray}
V &=& m_1^2 \Phi^\dagger \Phi + m_2^2 \eta^\dagger \eta + {1 \over 2} 
\lambda_1 (\Phi^\dagger \Phi)^2 + {1 \over 2} \lambda_2 (\eta^\dagger \eta)^2 
\nonumber \\ &+& \lambda_3 (\Phi^\dagger \Phi)(\eta^\dagger \eta) + 
\lambda_4 (\Phi^\dagger \eta)(\eta^\dagger \Phi) + \mu_{12}^2 (\Phi^\dagger 
\eta + \eta^\dagger \Phi),
\end{eqnarray}
where the $\mu_{12}^2$ term breaks lepton number softly and is the only 
possible such term.  Let $\langle \phi^0 \rangle = v$, $\langle \eta^0 \rangle 
= u$, then the equations of constraint for the minimum of $V$ are given by
\begin{eqnarray}
&& v[m_1^2 + \lambda_1 v^2 + (\lambda_3 + \lambda_4) u^2] + \mu_{12}^2 u = 0, 
\\ && u[m_2^2 + \lambda_2 u^2 + (\lambda_3 + \lambda_4) v^2] + \mu_{12}^2 v 
= 0.
\end{eqnarray}
Consider the case $m_1^2 < 0$, $m_2^2 > 0$, and $|\mu_{12}^2| << m_2^2$, then
\begin{equation}
v^2 \simeq -{m_1^2 \over \lambda_1}, ~~~ u \simeq -{\mu_{12}^2 v \over 
m_2^2 + (\lambda_3 + \lambda_4) v^2}.
\end{equation}
Hence $u$ may be very small compared to $v$(= 174 GeV).  For example, if 
$m_2 \sim 1$ TeV, $|\mu_{12}^2| \sim 10$ GeV$^2$, then $u \sim 1$ MeV and
\begin{equation}
m_\nu \sim \left( {f \over 1.0} \right)^2 \left( {1~{\rm TeV} \over m_N} 
\right) ~{\rm eV}.
\end{equation}

Since both $m_N$ and $m_2$ are now of order 1 TeV, they may be produced 
at future colliders and be detected.  (I) If $m_2 > m_N$, then the physical 
charged Higgs boson $h^+$, which is mostly $\eta^+$, will decay into $N$, 
which then decays into a charged lepton and a $W$ boson via $\nu - N$ miixing:
\begin{equation}
h^+ \to l_i^+ N_j, ~~~ N_j \to l_k^\pm W^\mp.
\end{equation}
(II) If $m_N > m_2$, then
\begin{equation}
N_i \to l_j^\pm h^\mp, ~~~ h^+ \to t \bar b,
\end{equation}
the latter coming from $\Phi - \eta$ mixing.  In either case, $m_2$ and $m_N$ 
can be determined kinematically, and $|f_{ij}|$ measured up to an overall 
scale.

In summary, the particle spectrum of the leptonic Higgs doublet model 
consists of the usual Standard-Model particles, including the one physical 
Higgs boson $h_1^0$, 3 heavy $N_R$'s at the TeV scale, and a heavy scalar 
doublet $(h^\pm, h_2^0, A)$ of individual masses $\sim m_2$.  The charged 
Higgs boson $h^\pm$ can be pair-produced at hadron colliders, whereas $N_R$ 
($h^\pm$) can be produced at lepton colliders via the exchange of 
$h^\pm$ ($N_R$).

\section{The Size of Lepton Number Violation}

It has been shown in the above that whereas Majorana neutrino masses have to 
be tiny, the actual magnitude of lepton number violation may come in all 
sizes.

(1) \underline {Large}: $m_N \sim 10^{13}$ GeV in the canonical seesaw 
mechanism.

(2) \underline {Medium}: $|\mu_{12}^2| \sim 10$ GeV$^2$ in the leptonic 
Higgs doublet model with $m_N \sim 1$ TeV.

(3) \underline{Small}: $\lambda_0 \langle \chi \rangle \sim 10$ eV in the 
Higgs triplet model ($m_\xi \sim 1$ TeV) with a singlet bulk scalar in extra 
large dimensions.

In (2) and (3), direct experimental determination of the relative magnitudes 
of the elements of ${\cal M}_\nu$ is possible at future colliders.

\section{Muon Anomalous Magnetic Moment}

The recent measurement \cite{g-2} of the muon anomalous magnetic moment 
appears to disagree with the Standard-Model prediction \cite{cm} by 
2.6$\sigma$, i.e.
\begin{equation}
\Delta a_\mu = a_\mu^{exp} - a_\mu^{SM} > 215 \times 10^{-11}
\end{equation}
at 90\% confidence level. The origin of this discrepancy may be directly 
related to the TeV physics responsible for neutrino mass.  In the leptonic 
Higgs doublet model, this has the following consequences. \cite{mr}

Assume all $m_N$'s are equal with its Yukawa coupling matrix given by
\begin{equation}
h_{ij} = \left[ \begin{array} {c@{\quad}c@{\quad}c} 2 c h_1 & -\sqrt 2 s h_1 & 
\sqrt 2 s h_1 \\ 2 s h_2 & \sqrt 2 c h_2 & -\sqrt 2 c h_2 \\ 0 & \sqrt 2 h_3 & 
\sqrt 2 h_3 \end{array} \right],
\end{equation}
with $h_1 \leq h_2 \leq h_3$ and $s = \sin \theta$, $c = \cos \theta$, we 
then find
\begin{equation}
m_\eta < 371 \sqrt {\alpha_h} ~{\rm GeV},
\end{equation}
where $h_1 \simeq h_2 \simeq h_3$ has been assumed.  In other words, the 
neutrino mass matrix has nearly degenerate mass eigenvalues.  If not, then 
satisfying Eq.~(15) would be in conflict with the experimental upper limit 
on the $\tau \to \mu \gamma$ branching fraction.  As it is, we also have 
the interesting prediction of 
\begin{eqnarray}
&& {\Gamma (\mu \to e \gamma) \over m_\mu^5} ~:~ {\Gamma (\tau \to e \gamma) 
\over m_\tau^5} ~:~ {\Gamma (\tau \to \mu \gamma) \over m_\tau^5} \nonumber 
\\ &=& 2 s^2 c^2 (\Delta m^2)^2_{sol} ~:~ 2 s^2 c^2 (\Delta m^2)^2_{sol} 
~:~ (\Delta m^2)^2_{atm}
\end{eqnarray}

In Fig.~(1), the branching fractions of $\tau \to \mu \gamma$ and $\mu \to 
e \gamma$, and the $\mu - e$ conversion ratio in $^{13}Al$ are plotted 
using the lower bound of Eq.~(15), as a function of the common neutrino 
mass $m_\nu$.  Using Eq.~(16), the values
\begin{equation}
(\Delta m^2)_{atm} = 3 \times 10^{-3} ~{\rm eV}^2, ~~~ (\Delta m^2)_{sol} = 
3 \times 10^{-5} ~{\rm eV}^2,
\end{equation}
have been chosen according to present data from neutrino-oscillation 
experiments.  At $m_\nu \simeq 0.2$ eV, which is in the range of present 
upper limits on $m_\nu$ from neutrinoless double beta decay, $B(\mu \to e 
\gamma)$ and $R_{\mu e}$ are both at their present experimental upper limits. 
Hence Eq.~(18) will be tested in new experiments planned for the near 
future which will lower these upper limits.

\section*{Acknowledgements}
I thank Milla Baldo Ceolin and the other organizers of the 9th International 
Workshop on ``Neutrino Telescopes'' for their great hospitality and a 
stimulating conference.  This work was supported in part by the 
U.~S.~Department of Energy under Grant No.~DE-FG03-94ER40837.

\begin{figure}
\vspace*{13pt}
         \mbox{\epsfig{figure=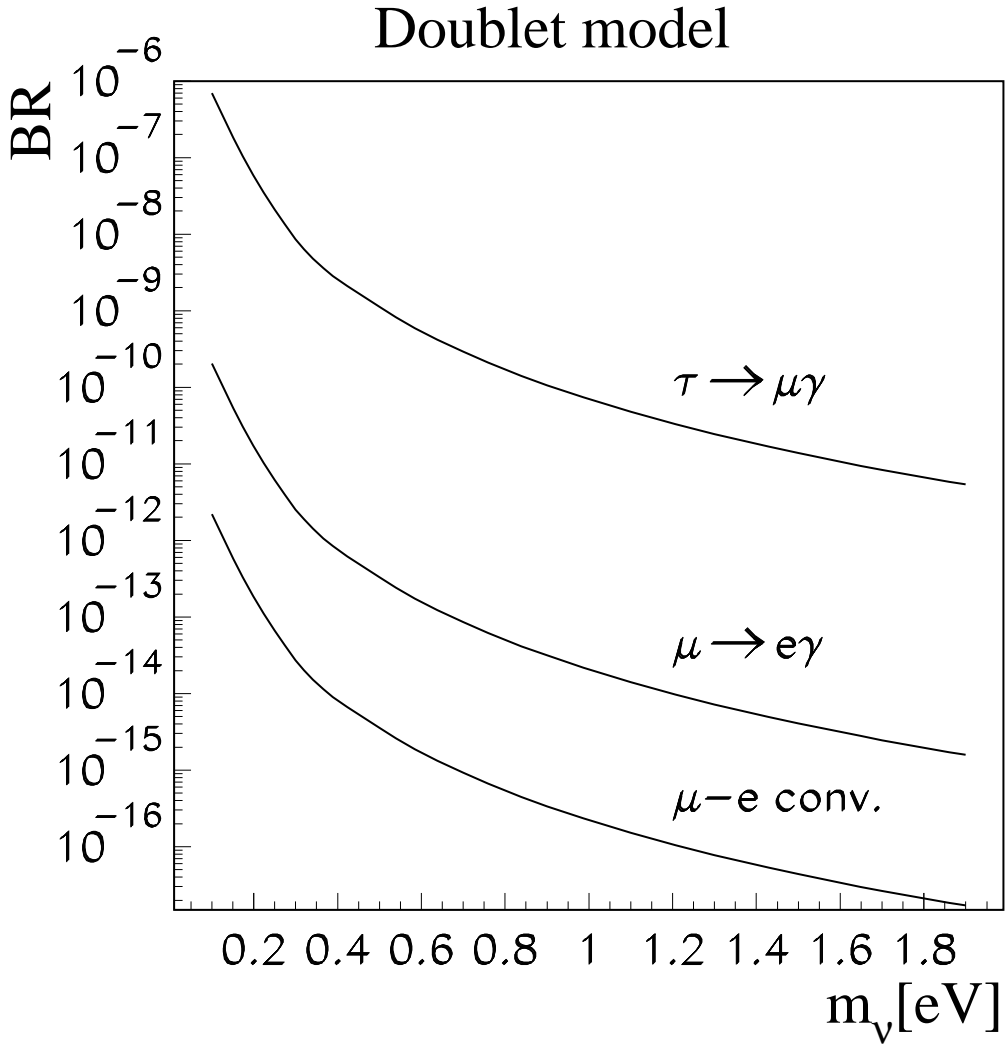,width=15.0cm}}
\caption{Lower bounds on $B(\tau \to \mu \gamma)$, $B(\mu \to e \gamma)$, and 
$R_{\mu e}$ \protect.}
\end{figure}

\end{document}